\documentclass[preprint t]{elsarticle}

\usepackage{lineno,hyperref}
\modulolinenumbers[5]

\usepackage{amsfonts,bbm,amssymb}
\usepackage{dsfont}
\usepackage{txfonts}
\usepackage{hyperref}
\usepackage[toc,page]{appendix}

\def\bra#1{\mathinner{\langle{#1}|}}
\def\ket#1{\mathinner{|{#1}\rangle}}

\newcommand{\ii}{ {\rm i} }
\newcommand{\dd}{ {\rm d} }
\newcommand{\TT}{\mathbb{T}}
\newcommand{\VV}{\mathbb{V}}
\newcommand{\ZZ}{\mathbb{Z}}
\newcommand{\RaR}{\mathbb{R}}
\newcommand{\CC}{\mathbb{C}}
\newcommand{\y}{{\rm y}}
\newcommand{\x}{{\rm x}}
\newcommand{\z}{{\rm z}}

\newcommand{\half}{\frac{1}{2}}
\newcommand{\eref}[1]{(\ref{#1})}
\newcommand{\mm}[1]{{\mathbf{#1}}}
\def\tr{{{\rm tr}}}

\def\one{\mathbbm{1}}
\def\Re{{\,{\rm Re}\,}}

\journal{Nuclear Physics B}

\begin{document}

\begin{frontmatter}
 
\title{Quasilocal conservation laws in $XXZ$ spin-$1/2$ chains:\\ 
open, periodic and twisted boundary conditions}

\author{Toma\v{z} Prosen}

\address{Department of Physics, Faculty of Mathematics and Physics, University of Ljubljana, Jadranska 19, SI-1000 Ljubljana, Slovenia}

\begin{abstract}
A continuous family of quasilocal exact conservation laws is constructed in the anisotro\-pic Heisenberg ($XXZ$) spin-$1/2$ chain for periodic (or twisted) boundary conditions and for a set of commensurate anisotropies densely covering the entire easy plane interaction regime.
All local conserved operators follow from the standard ({\em Hermitian}) transfer operator in fundamental representation (with auxiliary spin $s=1/2$), and are all even with respect to a spin flip operation. However, the quasilocal family is generated by differentiation of a {\em non-Hermitian highest weight transfer operator} with respect to a complex auxiliary spin representation parameter $s$ and includes also operators of odd parity. 
 For a finite chain with open boundaries the time derivatives of quasilocal operators are not strictly vanishing but result in operators localized near the boundaries of the chain.
We show that a simple modification of the non-Hermitian transfer operator results in exactly conserved, but still quasilocal operators for periodic or generally 
twisted boundary conditions. As an application, we demonstrate that implementing the new exactly conserved operator family for estimating the high-temperature spin Drude weight results, in the thermodynamic limit, in exactly the same lower bound as for almost conserved family and open boundaries.
Under the assumption that the bound is saturating (suggested by agreement with previous thermodynamic Bethe ansatz calculations) we propose a simple explicit construction of infinite time averages
of local operators such as the spin current.
\end{abstract}

\end{frontmatter}

\section{Introduction}
\label{intro}
The anisotropic Heisenberg spin $1/2$ chain, or the so-called $XXZ$ model, is probably the best studied quantum many body model with strong interactions.
This is mainly due to the fact that, on one hand, it provides a paradigmatic example of a completely integrable system for which computation of the complete energy spectrum and the corresponding eigenstates can be reduced to solving a system of coupled algebraic equations, the so-called Bethe equations, while on the other hand it can be used to describe the physics of magnetism in quasi one-dimensional solids, the so-called spin chain materials \cite{spinchains}. Many simple (say local) physical observables, as well as correlation functions, at temperature zero or at thermal equilibrium are thus amenable to explicit evaluation \cite{korepin,sklyanin,faddeev}.
Nevertheless, time-dependent phenomena and temporal-correlation functions, or other observables characterizing model's nonequilibrium or transport 
properties \cite{fabian} remain much harder to evaluate analytically \cite{essler} or even approximately, often involving unverifiable assumptions. A prime example of this kind has been the problem of spin Drude weight at finite temperatures \cite{castella,zotos,zotos2,carmelo,fabian2,benz,affleck,robin,karrasch,karrasch2} which raised controversies over several decades since various approximate or numerical approaches were yielding conflicting results. This issue has only recently been resolved \cite{prl11a,prl13} by proposing new quasilocal (almost) conserved quantities which lie outside  the scope of the traditional algebraic Bethe ansatz method. However, these new quantities, which derived from exact {\em steady state} solutions of boundary driven quantum master equations for the open chain \cite{prl11a,prl11b,kps,pks,iz,prl14,npb}, are not exactly conserved, but their time derivative amounts to terms localized at the chain boundaries. These steady states in turn can be related \cite{prl13} to infinitely dimensional solutions of the Yang-Baxter equation (or highest weight representations of the quantum group $U_q(\mathfrak{sl}_2)$ at complex value of spin representation parameter). The application of such {\em almost conserved} quasilocal operators to rigorous estimation of Drude weights is associated with nontrivial mathematical issues \cite{ip13} at finite (non-infinite) temperatures.

It is therefore highly desirable to clarify a possible existence of analogous quasilocal objects for periodic boundary conditions which would exactly commute with the Hamiltonian. This is what we achieve in the present work: by generalizing and slightly modifying the approach of Ref.~\cite{prl13} we explicitly construct holomorphic families of exactly conserved quasilocal operators for periodic as well as generally twisted boundary conditions. Half of these new operators are
odd with respect to spin flip symmetry and these remain orthogonal to all local conserved operators of algebraic Bethe ansatz. This paper also provides a fully rigorous background which justifies some details of a calculation reported in Ref.~\cite{prl13}.

In the rest of this section we shall define the model with different boundary conditions treated in this work. In section \ref{bd} we define transfer operators of the $XXZ$ with respect to arbitrary complex spin representation of the quantum symmetry group and relate its $s$-derivative to the solution of the corresponding boundary
driven Lindblad equation for the open chain. In section \ref{ob} we then discuss algebraic properties of such objects together with precise definition of quasi- and pseudo-locality of extensive spin chain operators. In section \ref{pbc} the main technical trick of the paper is presented which allows the aforementioned construction to extend to
periodic boundary conditions. Quasilocality of the new conservation laws for both types of boundary conditions on the corresponding domain of the spectral parameter is then rigorously proven in section \ref{proof}. In section \ref{spinflip} characterization of traditional local conserved operator and new quasilocal ones is given in terms
of spin flip parity symmetry, which explains why the quasilocal quantities are of prime importance for nonequilibrium physics. In section \ref{twisted} we then show how
periodic boundary conditions case straightforwardly generalizes to twisted boundary condition with an arbitrary gauging phase. In section \ref{appl} we finally discuss
the most direct application of the new exactly conserved quantities for periodic boundaries for providing rigorous lower bounds on finite temperature
dynamical susceptibilities. In particular, we rederive Mazur-Suzuki's theorem \cite{mazur,suzuki} for the case of a continuous set of conserved operators, formulating the general bound in terms of a solution of complex Fredholm integral equation of the first kind. Under the assumption that the bound is saturating, the result gives also an explicit expression for the time-averaged physical operator in terms of a quasilocal conserved set. Explicit results for the case of spin current and spin Drude weights are given for illustration.

\subsection{The $XXZ$ model}

We consider a chain of $n$ quantum spins $1/2$, described by Pauli matrices $\sigma^\alpha, \alpha\in\{\x,\y,\z,\pm,0\}$, $\sigma^\pm \equiv \half( \sigma^\x \pm \ii\sigma^\y )$, $\sigma^0\equiv \one_2$.
Here and below $\one_{d}$ denotes $d\times d$ unit matrix. 
Considering a local interaction over a pair of sites of the anisotropic Heisenberg form
\begin{equation}
h = 2\sigma^+ \otimes \sigma^- + 2\sigma^-\otimes \sigma^+ + \Delta \sigma^\z \otimes \sigma^\z,
\end{equation}
one defines the $XXZ$ Hamiltonian with trivial open boundaries (with no boundary fields) as a $2^n \times 2^n$ matrix, or an operator over the physical spin Hilbert space ${\cal H}^{\otimes n}_{\rm p}$, where ${\cal H}_{\rm p}\equiv\CC^2$,
\begin{equation}
H_{\rm obc} = \sum_{x=0}^{n-2} \one_{2^x} \otimes h \otimes \one_{2^{n-x-2}}.
\end{equation}
Similarly, one may introduce a $XXZ$ Hamiltonian with arbitrary twisted boundary condition by introducing a flux (phase) $\phi \in [0,2\pi)$:
\begin{equation}
H_{\phi} = H_{\rm obc} + 2e^{\ii\phi} \sigma^+ \otimes \one_{2^{n-2}} \otimes \sigma^- +2 e^{-\ii\phi} \sigma^-\otimes \one_{2^{n-2}} \otimes \sigma^+ + \Delta \sigma^\z \otimes \one_{2^{n-2}} \otimes \sigma^\z.
\label{Hphi}
\end{equation}
Note that for $\phi=0$ one obtains the more commonly studied $XXZ$ Hamiltonian with {\em periodic boundary conditions} $H_{\rm pbc} = H_0$. 
Using a unitary (canonical) transformation
\begin{equation}
C_\phi =\exp\left(\ii\frac{\phi}{n}\sum_{x=0}^{n-1} x\,\one_{2^{x}}\otimes\frac{\sigma^\z}{2}\otimes\one_{2^{n-1-x}}\right)
\label{canonical}
\end{equation}
the twisted Hamiltonian becomes manifestly periodic, i.e., it can be written in a $\ZZ_n$ translationally invariant form
\begin{equation}
H'_\phi=C_\phi H_\phi C^\dagger_\phi = \sum_{x=0}^{n-1} (2e^{-\ii \phi/n} \sigma^+_x \sigma^-_{x+1} +  2e^{\ii \phi/n} \sigma^-_x \sigma^+_{x+1} + \Delta \sigma^\z_x \sigma^\z_{x+1}),
\label{Hphi2}
\end{equation} 
if local spin variables are written as 
\begin{equation}
\sigma^\alpha_x = \one_{2^{x}} \otimes \sigma^\alpha \otimes \one_{2^{n-x-1}}
\end{equation}
and $x+1$ is taken ${\rm mod}\, n$. In the following we will only discuss the easy plane regime $|\Delta| \le 1$ where we parametrize the anisotropy as 
$\Delta = \cos \eta$, for  $\eta \in [0,\pi]$.

\section{Boundary driven chain and the nonequilibrium quantum transfer operator}
\label{bd}

$XXZ$ chain is intimately connected to the quantum group $U_q(\mathfrak{sl}_2)$ symmetry \cite{pasquier}, with $q=e^{\ii\eta}$,
whose generators $\mm{S}^\pm,\mm{S}^\z$ satisfy the $q-$deformed $\mathfrak{sl}_2$ algebra
\begin{equation}
[\mm{S}^+,\mm{S}^-]= \frac{\sin (2\eta \mm{S}^\z)}{\sin\eta},\quad [\mm{S}^\z,\mm{S}^\pm]=\pm \mm{S}^\pm.
\label{Uq}
\end{equation}
Here we shall facilitate its general ({\em non-unitary}) highest weight representation, para\-metrized by a complex parameter $s\in\CC$ (the so-called complex spin). Given the highest-weight-state $\ket{0}$ such that $\mm{S}^+_s\ket{0}=0$, 
explicit representation (unique up to unitary transformations), over infinitely dimensional Hilbert space spanned by a orthonormal basis $\{ \ket{k}; k=0,1,2\ldots \}$ -- Verma module ${\cal V}_s$ -- reads
\begin{eqnarray}
\mm{S}^\z_s &=& \sum_{k=0}^\infty (s-k) \ket{k}\bra{k}, \nonumber\\
\mm{S}^+_s &=& \sum_{k=0}^\infty \frac{\sin(k+1)\eta}{\sin\eta} \ket{k}\bra{k+1}, \label{verma} \\
\mm{S}^-_s &=& \sum_{k=0}^\infty \frac{\sin(2s-k)\eta}{\sin\eta} \ket{k+1}\bra{k}. \nonumber
\end{eqnarray}
For a dense set of commensurate anisotropies $\eta = \pi l/m, l,m\in\ZZ^+$, ${\cal V}_s$ becomes {\em finite} $m-$dimensional (truncated linear span of states ${\rm lsp}\{\ket{k};k\in\{0,1\ldots m-1\}\}$ as the states $\ket{m-1}$ and $\ket{m}$ are not connected by $\mm{S}^+_s$).
For $2s \in \ZZ^+$ (and any $\eta$), ${\cal V}_{s}$ becomes reducible to a well known $2s+1$ dimensional irrep, and {\em only} then the representation is {\em unitary}.
Moreover, only then the representation is {\em parity symmetric} in the sense that, for any $\eta$,
\begin{equation}
\mm{U} \mm{S}^\pm_s \mm{U}^{-1}=\mm{S}^\mp_s,\qquad \mm{U}\mm{S}^\z_s\mm{U}^{-1} = -\mm{S}^\z_s,\quad{\rm for}\quad 2s\in\ZZ^+
\label{aparity}
\end{equation}
where $\mm{U}\in {\rm End}({\cal V}_{s})$ is the spin-flip operation
\begin{equation}
\mm{U} = \sum_{k=0}^{2s} \ket{k}\bra{2s-k}.
\end{equation}

Quantum group $U_q(\mathfrak{sl}_2)$ defines the universal $R-$matrix $\mm{R}_{s,s'} (\varphi) \in {\rm End}({\cal V}_s\otimes {\cal V}_{s'})$ depending on {\em the spectral parameter} $\varphi$, as the solution of the Yang-Baxter equation (YBE) over a generic triple \cite{ttf83,k01,k02} ${\cal V}_s \otimes {\cal V}_{s'} \otimes {\cal V}_{s''}$ for arbitrary $s,s',s''\in \CC$. We consider the Lax operator as the $R-$matrix $\mm{R}_{s,1/2}$ having one leg in the physical spin space carying the {\em fundamental} representation $ {\cal V}_{1/2}\equiv {\cal H}_{\rm p}=\CC^2$ and the other one in the {\em auxiliary} space (so-called anzilla) $ {\cal V}_s \equiv {\cal H}_{\rm a}$,
 i.e. a $2\times 2$ matrix with entries\footnote{In our notation we use bold-upright letters to denote operators which are {\em not} scalars in auxiliary space.} in ${\rm End}({\cal V}_{s})$
\begin{equation}
\mm{L}(\varphi,s) = \pmatrix{
\sin(\varphi+\eta \mm{S}^\z_s) & (\sin\eta) \mm{S}^-_s \cr
(\sin\eta) \mm{S}^+_s & \sin(\varphi-\eta \mm{S}^\z_s) \cr
} = \sum_{\alpha \in{\cal J}} \mm{L}^\alpha(\varphi,s) \otimes \sigma^\alpha,
\end{equation}
where ${\cal J}=\{+,-,0,\z\}$ and
\begin{equation}
\mm{L}^0(\varphi,s) = \sin\varphi \cos(\eta \mm{S}^\z_s),\;\;
\mm{L}^\z(\varphi,s)  = \cos\varphi \sin(\eta \mm{S}^\z_s),\;\;
\mm{L}^\pm(\varphi,s) = (\sin \eta) \mm{S}_s^\mp.
\end{equation}
Then, the YBE over ${\cal V}_{s}\otimes {\cal V}_{s'}\otimes {\cal V}_{1/2}$ together with the fact that $\bra{0}\otimes\bra{0}$ ($\ket{0}\otimes\ket{0}$) is a left (right) eigenvector of the $R$-matrix over ${\cal V}_{s}\otimes {\cal V}_{s'}$
guarantees commutativity of the {\em highest-weight non-Hermitian transfer operator} (HNTO) \footnote{In order to avoid excessive use of indices and at the same time keep notation unambiguous  we adopt the following convention: For algebraic objects which are defined
as operators over tensor products over two or more different spaces ${\cal H}_{v}\otimes {\cal H}_{\rm other}$, say 
$v\in\{{\rm p},{\rm a}\}$, the symbol $\otimes_{v}$ will denote a {\em partial tensor product} with
respect to a space ${\cal H}_{v}$, making the resulting object acting over ${\cal H}_v\otimes {\cal H}_v\otimes {\cal H}_{\rm other}$,
and the usual operator (matrix) product with respect to all other spaces.
Concretely, writing $A = \sum_\mu a_\mu X_\mu$ and $B = \sum_\mu b_\mu Y_\mu$, where $a_\mu, b_\mu \in {\rm End}({\cal H}_\nu)$, $X_\mu,Y_\mu \in {\rm End}({\cal H}_{\rm other})$, one has 
 $A \otimes_\nu B = \sum_{\mu,\mu'} (a_\mu \otimes b_{\mu'}) X_{\mu} Y_{\mu'}$.
}
 $W_n(\varphi,s)\in {\rm End}({\cal H}^{\otimes n}_{\rm p})$ \cite{pip13}
\begin{equation}
W_n(\varphi,s) = \bra{0} \mm{L}(\varphi,s)^{\otimes_{\rm p} n} \ket{0}.
\label{HNTO}
\end{equation}
Namely, for any pair of {\em spectral} parameters $\varphi,\varphi'\in\CC$ and {\em representation} parameters $s,s'\in\CC$, we have
\begin{equation}
[W_n(\varphi,s),W_n(\varphi',s')] = 0.
\label{eq:com}
\end{equation}
The highest weight nature of the representation (\ref{verma}) immediately implies that the matrix $W_n(\varphi,s)$ is {\em lower triangular}. We note that the expression (\ref{HNTO}) generates a matrix product
operator (MPO) representation of HNTO
\begin{equation}
W_n(\varphi,s) = \sum_{\alpha_1\ldots \alpha_n\in{\cal J}} \bra{0}\mm{L}^{\alpha_1} \mm{L}^{\alpha_2}\cdots\mm{L}^{\alpha_n}\ket{0} \sigma^{\alpha_1}\otimes \sigma^{\alpha_2}\cdots \otimes \sigma^{\alpha_n}
\end{equation}

On the other hand, considering YBE over ${\cal V}_{1/2}\otimes {\cal V}_{1/2} \otimes {\cal V}_s$ and the fact that hamiltonian density can be generated as $\partial_\varphi \mm{R}_{1/2,1/2}(\varphi) |_{\varphi=0} \propto h$ (see e.g. \cite{korepin}), one
obtains a fundamental divergence relation for local two-site commutators
 \cite{S70,sklyanin}
\begin{equation}
[h,\mm{L} \otimes_{\rm p} \mm{L}] = 2\sin\eta\, (\mm{L} \otimes_{\rm p} \mm{L}_\varphi - \mm{L}_\varphi\otimes_{\rm p}\mm{L} ),
\label{divergence}
\end{equation}
where $\mm{L}\equiv\mm{L}(\varphi,s)$, $\mm{L}_\varphi\equiv \partial_\varphi\mm{L}(\varphi,s) = \cos\varphi \cos(\eta\mm{S}^\z_s)\otimes\sigma^0-\sin\varphi\sin(\eta\mm{S}^\z_s)\otimes\sigma^\z$.
Left-tensor-multiplying Eq.~\eref{divergence} by $\bra{0} \mm{L}^{\otimes_{\rm p} (x-1)}$, right-tensor multiplying it by $\mm{L}^{\otimes_{\rm p} (n-x)}\ket{0}$, and summing over $x\in\{1\ldots n\}$, we obtain a useful identity
\begin{equation}
[H_{\rm obc},W_n(\varphi,s)] = -\tau\otimes W_{n-1}(\varphi,s) + W_{n-1}(\varphi,s) \otimes \tau,
\label{HNTOrel}
\end{equation}
where $\tau$ is a diagonal $2\times 2$ matrix
\begin{equation}
\tau = 2\sin\eta\left[ (\cos\varphi \cos\eta s)\,\sigma^0 - (\sin\varphi \sin\eta s)\,\sigma^\z \right].
\end{equation}

It has been shown in Ref.~\cite{prl11b} that if $2^n \times 2^n$ {\em upper triangular matrix} $S_{\!n}$ with {\em unit} diagonal elements satisfies the {\em defining relation}
\begin{equation}
[H_{\rm obc},S_{\!n}] = -\ii \varepsilon \left( \sigma^\z \otimes S_{\!n-1} - S_{\!n-1}\otimes \sigma^\z \right)
\label{defS}
\end{equation}
then
\begin{equation}
\rho_\infty = \frac{S_{\!n} S_{\!n}^\dagger}{\tr(S_{\!n} S_{\!n}^\dagger)}
\end{equation}
is the (unique) nonequilibrium steady state density operator of the maximally boundary driven Lindblad dynamics 
\begin{equation}
\frac{\dd}{\dd t}\rho_t = -\ii [H_{\rm obc},\rho_t] + \varepsilon \sum_{j=1}^2 \left(2 A_j \rho_t A_j^\dagger - \{A_j^\dagger A_j,\rho_t\}\right)
\label{lind}
\end{equation}
with a pair of ultra-local incoherent boundary-jump processes $A_1 = \sigma^+ \otimes \one_{2^{n-1}}$, $A_2 = \one_{2^{n-1}}\otimes \sigma^-$, with the rates $\epsilon$.  
Comparing the relations (\ref{HNTOrel}) and (\ref{defS}) one may identify HNTO with the fixed point of Lindblad dynamics
in exactly two non-equivalent cases: (i) For $\varphi=0$ one finds
\begin{equation}
S_{\!n} = W_n^T(0,s) \frac{(\sigma^\z)^{\otimes n}}{(\sin\eta s)^n}\quad {\rm for} \quad \cot \eta s = -\frac{\ii\varepsilon}{2\sin \eta},
\end{equation}
while (ii) for $\varphi=\pi/2$ one finds
\begin{equation}
S_{\!n} = W_n^T\left(\frac{\pi}{2},s\right) \frac{1}{(\cos\eta s)^n}  \quad {\rm for} \quad \tan \eta s = \frac{\ii\varepsilon}{2\sin \eta}.
\end{equation}
In both cases, the steady state solution of boundary-driven nonequilibrium problem (\ref{lind}) requires imaginary spin $s\in\ii \RaR$ representation which is therefore always nonunitary.
Since with fixed diagonal of $S_{\!n}$ the Cholesky decomposition $S_{\!n} S_{\!n}^\dagger$ is unique, one thereby also obtains an interesting symmetry relation for HNTO
\begin{equation}
W_n\left(\frac{\pi}{2},s\right) = (-\sigma^\z)^{\otimes n} W_n\left( 0, s+ \frac{\pi}{2\eta}\right).
\end{equation}
Another remarkable property of HNTO is the spin-inversion identity
\begin{equation}
W_n(\varphi,s) W_n(\varphi,-s) = (\sin(\varphi+\eta s)\sin(\varphi-\eta s))^n \one_{2^n},
\end{equation}
which can be proved straightforwardly by writing the LHS as an iterative map over ${\cal H}_{\rm a}\otimes {\cal H}_{\rm a}$, sandwiched between
$\bra{0}\otimes\bra{0}$ and $\ket{0}\otimes \ket{0}$, and showing that
all matrix elements in physical space ${\cal H}^{\otimes n}_{\rm p}$ should vanish except for trivial diagonal ones.

\section{Quasilocal almost conserved operator family for open boundaries}
\label{ob}

HNTO (\ref{HNTO}) is neither a local operator, nor it is conserved in time as its time derivative (\ref{HNTOrel}) is a non-local object. Yet, it can be used to generate a very interesting family of operators in terms of differentiation with respect to the spin representation parameter $s$ around the {\em scalar point} $s=0$
\begin{equation}
Z_n(\varphi) = \frac{1}{2(\sin\varphi)^{n-2}\eta\sin\eta}\partial_s W_n(\varphi,s)\vert_{s=0} - \frac{\sin\varphi\cos\varphi}{2\sin\eta} M^\z_n,
\label{Zdef}
\end{equation}
where $M^\z_n=\sum_{x=0}^{n-1}\one_{2^x}\otimes\sigma^\z\otimes\one_{2^{n-1-x}}$ is the conserved component of magnetization.
The $s-$derivative can be implemented as an MPO in terms of an additional `derivative anzilla' qubit ${\cal H}_{\rm b}=\CC^2$, 
\begin{equation}
Z_n(\varphi) = \frac{\sin^2\varphi}{2\eta\sin\eta}\bra{0}_{\rm a}\bra{0}_{\rm b}\tilde{\mm{L}}(\varphi)^{\otimes_{\rm p} n} \ket{0}_{\rm a}\ket{1}_{\rm b} - 
\frac{\sin\varphi\cos\varphi}{2\sin\eta} M^\z_n,
\end{equation}
defining an {\em extended Lax operator} $\tilde{\mm{L}}(\varphi) \in {\rm End}({\cal H}_{\rm a}\otimes {\cal H}_{\rm b}\otimes {\cal H}_{\rm p})$
\begin{equation}
\mm{\tilde{L}}(\varphi) = \frac{1}{\sin\varphi}\pmatrix{
\mm{L}(\varphi,0) & \partial_s \mm{L}(\varphi,s)|_{s=0} \cr
0 & \mm{L}(\varphi,0)} =\mm{L}_0(\varphi) \one_{\rm b} + \mm{L}_1(\varphi)\sigma^+_{\rm b},
\end{equation}
where 
\begin{equation}
\mm{L}_0(\varphi):= (\csc\varphi)\mm{L}(\varphi,0),\quad
\mm{L}_1(\varphi):= (\csc\varphi)\partial_s\mm{L}(\varphi,s)|_{s=0}.
\label{L01def}
\end{equation}
We shall refer to the operator family $Z_n(\varphi)$ as the {\em modified} highest-weight non-Hermitian transfer operators (mHNTO). 
It can be shown that $Z_n(\varphi)$ are quasilocal operators whose time-derivative is localized at the chain boundaries for a suitable domain $\varphi\in {\cal D}\subset\CC$.
Indeed, differentiating \eref{HNTOrel} w.r.t. $s$ at $s=0$ and using the definition we immediately obtain a very insightful relation
\begin{equation}
[H_{\rm obc},Z_n(\varphi)] =  \sigma^\z\otimes\one_{2^{n-1}} - \one_{2^{n-1}}\otimes\sigma^\z - 
2\sin\eta\cot\varphi\left( \sigma^0\otimes Z_{n-1}(\varphi)-Z_{n-1}(\varphi)\otimes\sigma^0\right). \label{HZc}
\end{equation}
Writing the Lax operator components $\tilde{\mm{L}}^\alpha \in {\rm End}({\cal H}_{\rm a}\otimes {\cal H}_{\rm b})$, $\mm{L}^\alpha \in {\rm End}({\cal H}_{\rm a})$, 
via $\tilde{\mm{L}}(\varphi) = \sum_{\alpha\in{\cal J}} \tilde{\mm{L}}^\alpha(\varphi) \otimes \sigma^\alpha$, 
$\tilde{\mm{L}}^\alpha(\varphi) = \mm{L}^\alpha_0(\varphi) \one_{\rm b} + \mm{L}^\alpha_1(\varphi)\sigma^+_{\rm b} $
satisfying the following {\em boundary transition conditions}
\begin{eqnarray}
\bra{0}_{\rm a}\bra{0}_{\rm b} \tilde{\mm{L}}^0 &=& \bra{0}_{\rm a}\bra{0}_{\rm b}, \quad \bra{0}_{\rm a}\bra{0}_{\rm b} \tilde{\mm{L}}^+ = 0, \nonumber \\
\tilde{\mm{L}}^0 \ket{0}_{\rm a}\ket{1}_{\rm b} &=&  \ket{0}_{\rm a}\ket{1}_{\rm b}, \quad \tilde{\mm{L}}^- \ket{0}_{\rm a}\ket{1}_{\rm b} =  0, \nonumber \\
\quad\tilde{\mm{L}}^\z \ket{0}_{\rm a}\ket{1}_{\rm b} &=&  \eta\cot\varphi \ket{0}_{\rm a} \ket{0}_{\rm b}, \quad\tilde{\mm{L}}^{\z,\pm} \ket{0}_{\rm a}\ket{0}_{\rm b} =  0, 
\label{bc}
\end{eqnarray}
one sees that mHNTOs allow for an expression in terms of open boundary translationally invariant sum of local operators
\begin{equation}
Z_n(\varphi) = \sum_{r=2}^n \sum_{x=0}^{n-r} \one_{2^{x}} \otimes q_r(\varphi) \otimes \one_{2^{n-r-x}},
\label{Zloc}
\label{qm}
\end{equation}
where $q_r(\varphi) \in {\rm End}({\cal H}^{\otimes r}_{\rm p})$ are local $r-$point operator densities with MPO representation:
\begin{equation}
q_r(\varphi) = \frac{\sin^2\varphi}{2\eta\sin\eta}\sum_{\alpha_2\ldots \alpha_{r-1}\in{\cal J}}\!\!\bra{0}_{\rm a}\bra{0}_{\rm b} \tilde{\mm{L}}^- \tilde{\mm{L}}^{\alpha_2}\cdots\tilde{\mm{L}}^{\alpha_{r-1}}\tilde{\mm{L}}^+ \ket{0}_{\rm a}\ket{1}_{\rm b} \sigma^- \otimes \sigma^{\alpha_2}\cdots \sigma^{\alpha_{r-1}}\otimes \sigma^+,
\label{qr}
\end{equation}
while the $r=2$ case has to be given separately, $q_2(\varphi) = \sigma^- \otimes \sigma^+$.
Note that $r=1$ term is exactly cancelled by the magnetization term subtracted in the definition (\ref{Zdef}).
Alternatively, since $\mm{L}^+_0(\varphi)\ket{0}=0$, the $s-$derivative should always hit the last factor and one may also write more explicitly (and usefully)
\begin{eqnarray}
q_r(\varphi)\!&=&\!\frac{\sin^2\varphi}{2\eta\sin\eta}\!\sum_{\alpha_2\ldots \alpha_{r-1}\in{\cal J}}\!\!\!\bra{0}\mm{L}^-_0(\varphi) \mm{L}^{\alpha_2}_0(\varphi)\cdots\mm{L}^{\alpha_{r-1}}_0(\varphi)\mm{L}^+_1(\varphi) \ket{0}  \sigma^- \otimes \sigma^{\alpha_2}\cdots \sigma^{\alpha_{r-1}}\otimes \sigma^+ \nonumber \\
&=&\!\sum_{\alpha_2\ldots \alpha_{r-1}\in{\cal J}} \bra{1}\mm{L}^{\alpha_2}_0(\varphi)\cdots\mm{L}^{\alpha_{r-1}}_0(\varphi)\ket{1}  \sigma^- \otimes \sigma^{\alpha_2}\cdots \sigma^{\alpha_{r-1}}\otimes \sigma^+.
\label{qr2}
\end{eqnarray}

Using the local operator sum ansatz (\ref{Zloc}) one is able to rewrite the RHS of (\ref{HZc}) in a form of a sum of operators localized at the boundaries
\begin{eqnarray}
[H_{\rm obc},Z_n(\varphi)] &=& \sigma^\z\otimes \one_{2^{n-1}} - \one_{2^{n-1}}\otimes\sigma^\z \nonumber \\ 
&+&  2\sin\eta\cot\varphi \sum_{r=2}^n \left(q_r(\varphi)\otimes \one_{2^{n-r}} - \one_{2^{n-r}}\otimes q_r(\varphi) \right).
\end{eqnarray} 
In the rest of this paper we show that there are important parameter regimes for which the operator sequence $\{ q_r(\varphi); r=2,3\ldots\}$ is quickly decreasing in a suitable operator norm, so the operator family (\ref{Zloc}) can be considered as {\em quasilocal} and {\em almost conserved}.

\newdefinition{deff}{Definition}
\begin{deff}
{\em Quasilocality:} An operator sequence $Z_n \in {\rm End}({\cal H}^{\otimes n}_{\rm p})$ which can be written as an open boundary translationally invariant sum of local operators $q_r$, like (\ref{Zloc}), for {\em any} $n$, is called {\em quasilocal} if there exist positive constants $\gamma,\xi > 0$, such that 
\begin{equation}
\| q_r \|_{\rm HS} \le \gamma e^{-\xi r},
\end{equation} 
where, for any matrix $a$,
\begin{equation}
\| a\|^2_{\rm HS} := \frac{\tr(a^\dagger a)}{\tr\one}
\end{equation} is a normalized Hilbert-Schmidt norm which satisfies a nice extensivity property
\begin{equation}
\| a \|_{\rm HS} = \|\,a \otimes \one_d \|_{\rm HS},\quad \forall d,
\label{extensivity}
\end{equation}
\end{deff}
as well as the normalized Cauchy-Schwartz inequality
\begin{equation}
\left |\frac{\tr (a b)}{\tr \one}\right| \le \| a\|_{\rm HS}\,\| b\|_{\rm HS}.
\label{CS}
\end{equation}
We remark that the Hilbert-Schmidt operator norm is the natural norm for high-tempe\-rature statistical mechanics as it is linked to an infinite temperature, tracial state $\omega_0(a) = \tr a/\tr\one$, namely $\|a\|^2_{\rm HS} = \omega_0(a^\dagger a)$.
Note also that it satisfies a useful inequality in relation to a $C^*$ operator norm $\| b \|^2 := \sup_{\omega} \omega(b^\dagger b)$, namely for any pair of bounded operators $a,b$ (say, elements of ${\rm End}({\cal H}^{\otimes n}_{\rm p})$), $\|a b\|_{\rm HS} \le \| a\|_{\rm HS} \| b \|$.

It is important to note also that the definition of quasilocality here differs from the standard one in $C^*$ statistical mechanics \cite{BR} which is based on the operator norm.
\begin{deff}
{\em Pseudolocality \cite{p99}:} An operator sequence $Z_n \in {\rm End}({\cal H}^{\otimes n}_{\rm p})$ of the form (\ref{Zloc}) is called {\em pseudolocal} if there exists a positive constant $K > 0$, such that 
\begin{equation}
\| Z_n\|^2_{\rm HS} \le K n.
\end{equation}
\end{deff}

Clearly, quasilocality implies pseudolocality as follows straightforwardly from the definitions\footnote{See e.g. end of section \ref{proof} for explicit demonstration.}. Pseudolocality is in fact the weakest definition of {\em spatial extensivity} of physical observables and to control it shall be of utmost importance for applications in nonequilibrium statistical mechanics, the example of which we shall discuss in section \ref{appl}.
We will show in the following sections that mHNTO $Z_n(\varphi)$ for $XXZ$ chain at any commensurate anisotropy $\eta = \pi l/m, l,m\in\ZZ^+$, and its extensions for periodic and twisted boundary conditions, are quasilocal operators in an appropriate domain of 
$\varphi$.

\section{Quasilocal conserved operator family for periodic boundary conditions}

\label{pbc}

So far, our constructions were meaningful for any value of anisotropy parameter $\eta$. From now on we shall restrict ourselves to the critical line $|\Delta|<1$ (easy plane anisotropy), and in particular, to a countable but {\em dense} set of {\em commensurate aniso\-tro\-pies}
\begin{equation}
\eta = \frac{\pi l}{m},\quad {\rm coprime\;} l,m\in\ZZ^+,\; m\neq 0,\;  l \le m.
\end{equation}
Under such condition, as discussed in section \ref{bd}, the auxiliary space becomes $m-$di\-mensional ${\cal H}_{\rm a}={\rm lsp}\{\ket{k};k=0,\ldots m-1\} \equiv {\cal V}_s$ for any value of complex spin $s$.
Then, one can define translationally invariant {\em periodic} non-Hermitian transfer operator (PNTO) in terms of a trace operation
\begin{equation}
V_n(\varphi,s) = \tr_{\rm a}\left\{\mm{L}(\varphi,s)^{\otimes_{\rm p} n}\right\}.
\label{Vdef}
\end{equation}
In analogy to HNTO, YBE in  ${\cal V}_{s}\otimes {\cal V}_{s'}\otimes {\cal V}_{1/2}$ and ${\cal V}_{1/2}\otimes {\cal V}_{1/2}\otimes {\cal V}_{s}$, immediately implies commutativity
\begin{equation}
[V_n(\varphi,s),V_n(\varphi',s')] = 0,\quad [H_{\rm pbc},V_n(\varphi,s)] = 0, \quad \forall s,s',\varphi,\varphi'.
\end{equation}
Similarly, YBE in  ${\cal V}_{s}\otimes {\cal V}^T_{s'}\otimes {\cal V}_{1/2}$ where ${\cal V}^T_{s'}$ is the {\em transposed} spin-$s'$ representation, implies
\begin{equation}
[V_n(\varphi,s),V_n^T(\varphi',s')] = 0.
\end{equation}
Note, however, since the transposed representation exchanges the roles of highest- and lowest-weight states, similar commutativity {\em does not} hold for the HNTOs, i.e., $[W_n(\varphi,s),W_n^T(\varphi',s')] \neq 0$.
Only in case $2s\in\ZZ^+$ the PNTO in fact becomes Hermitian\footnote{Strictly, it is Hermitian only for $\varphi\in\RaR$, when it is in fact even a {\em real symmetric matrix} in the standard basis where $(\sigma^\pm)^T=\sigma^\mp$.}
 $V_n^T(\varphi,s)\equiv V_n(\varphi,s)$. In fundamental representation $s=1/2$, $V(\varphi,1/2)$ is the standard transfer operator of {\em algebraic Bethe ansatz} \cite{korepin} and generates all the local conserved operators \cite{gm} $Q^{(j)}_n,j=1,2\ldots n-1$, such that $H_{\rm pbc}\propto Q_n^{(1)}$:
 \begin{equation}
 Q^{(j)}_n =  \partial^j_\varphi\log V_n(\varphi,1/2)|_{\varphi=\eta/2}.
 \end{equation}

\begin{figure}
        \centering	
	\includegraphics[width=7cm]{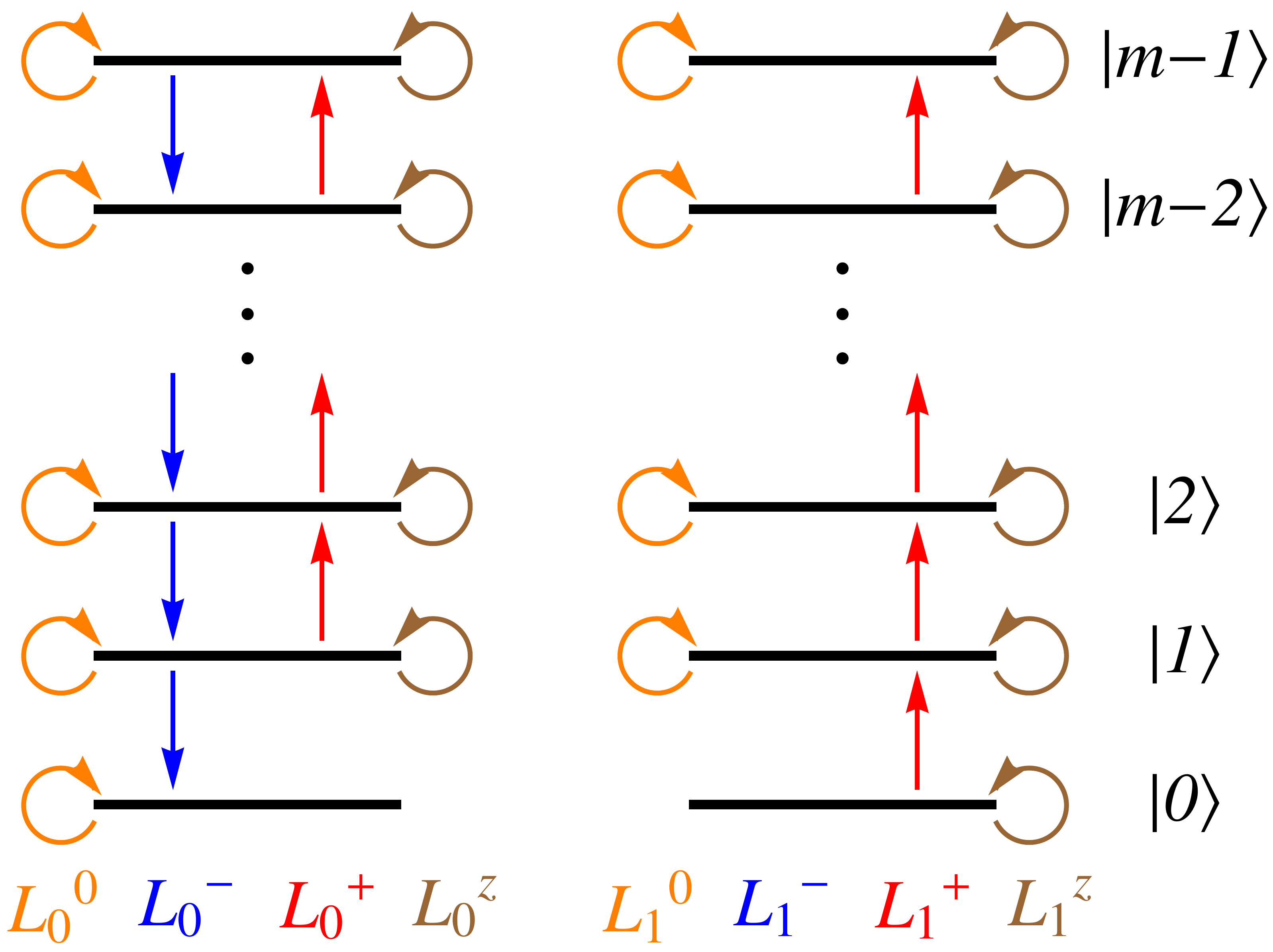}
	\caption{All allowed transitions -- indicated by arrows -- with the Lax operators (\ref{L01def}) [see Eq.~\eref{L01} for explicit
	transition amplitudes] among auxiliary basis states. The left tower shows the `regular' transitions $\mm{L}_0^\alpha$ (components $\alpha$ in different colors), 
	while the right tower shows possible `single defect' transitions $\mm{L}_1^\alpha$ due to $s$-differentiation. Note that the regular self-transition $\mm{L}_0^0$ 
	on the ground state $\ket{0}$ has an amplitude $1$ while all other regular self-transitions (by $\mm{L}_0^0$ or $\mm{L}_0^\z$) have amplitudes in modulus strictly less 
	than $1$ (for $\varphi \in {\cal D}_m$), which is the intuitive origin of quasilocality (proven in section \ref{proof}). Each term of the quasilocal operator density $q_r$ in the Pauli basis 
	(\ref{qr}) can be associated 
	with an $r-$step recurrent walk $\ket{0} \to \ket{0}$ with exactly one defect, therefore never visiting the state $\ket{0}$ in between. 
	Similarly, the terms of the remainder operators $p_n$ (\ref{p}) can be identified with 
	recurrent $n-$step walks starting and ending at the excited state $\ket{k}\to\ket{k},k > 0$, hence they can never visit the ground state in between. This in turn implies 
	exponential smallness (in $n$) of the norms of the remainder terms.}
	\label{Figure}
\end{figure}

Similarly as in the open boundary case, we define in the next step a family of {\em modified} periodic non-Hermitian transfer operators (mPNTO) by $s-$differentiation
\begin{eqnarray}
Y_n(\varphi) &=&   \frac{1}{2(\sin\varphi)^{n-2}\eta\sin\eta}\partial_s V_n(\varphi,s)\vert_{s=0}-\frac{\sin\varphi\cos\varphi}{2\sin\eta}  M^\z_n \nonumber \\
&=&  \frac{\sin^2\varphi}{2\eta\sin\eta}\tr_{\rm a}\left\{\bra{0}_{\rm b}\tilde{\mm{L}}(\varphi)^{\otimes_{\rm p} n} \ket{1}_{\rm b}\right\}- \frac{\sin\varphi\cos\varphi}{2\sin\eta} M^\z_n,
\label{Ydef}
\end{eqnarray}
which, clearly, again form a commuting and exactly conserved family
\begin{equation}
[Y_n(\varphi),Y_n(\varphi')] = 0,\quad [Y_n(\varphi),Y^T_n(\varphi')] = 0,\quad [H_{\rm pbc},Y_n(\varphi)] = 0,\quad\forall \varphi,\varphi'.
\label{cons}
\end{equation}
Let us define a periodic-left-shift as a linear map $\hat{\cal S} : {\rm End}({\cal H}^{\otimes n}_{\rm p}) \to {\rm End}({\cal H}^{\otimes n}_{\rm p})$ which is completely specified by its action on the Pauli basis
\begin{equation}
\hat{\cal S}(\sigma^{\alpha_0}\otimes \sigma^{\alpha_1} \otimes\cdots\sigma^{\alpha_{n-2}}\otimes\sigma^{\alpha_{n-1}}) = \sigma^{\alpha_{1}}\otimes \sigma^{\alpha_2}\otimes\cdots\sigma^{\alpha_{n-1}}\otimes\sigma^{\alpha_{0}}.
\end{equation}
Clearly, the definitions (\ref{Vdef},\ref{Ydef}) imply periodic-shift invariance of the PNTOs
\begin{equation}
\hat{\cal S} V_n(\varphi,s) = V_n(\varphi,s),\qquad
\hat{\cal S} Y_n(\varphi) = Y_n(\varphi).
\end{equation}
We shall now prove the following useful result which connects the modified non-Hermitian transfer operators for open and periodic boundary conditions:
\newdefinition{lemma}{Lemma}
\begin{lemma}
Using the operator densities (\ref{qr}) we find the following periodic translationally invariant expression for mPNTO
\begin{equation}
Y_n(\varphi) = \sum_{r=2}^n \sum_{x=0}^{n-1} \hat{\cal S}^x ( \one_{2^{n-r}}\otimes q_r(\varphi)) + \sum_{x=0}^{n-1}\hat{\cal S}^x (p_n(\varphi)),
\label{lem}
\end{equation}
where the `remainder' operator $p_n(\varphi)\in {\rm End}({\cal H}^{\otimes n}_{\rm p})$ is given as
\begin{equation}
p_n(\varphi) = \sum_{k=1}^{m-1} \bra{k}\mm{L}_0(\varphi)^{\otimes_{\rm p} (n-1)} \otimes_{\rm p} \mm{L}_1(\varphi)\ket{k}.
\label{p}
\end{equation}
\end{lemma}
\newproof{pf}{Proof}
\begin{pf}
The starting point is an obvious expression, following by applying the Leibniz rule to definition (\ref{Ydef}), then split into two terms:
\begin{eqnarray*}
 \frac{\partial_s V_n(\varphi,s)\vert_{s=0}}{(\sin\varphi)^{n}} &=& \sum_{x=0}^{n-1} \tr_{\rm a}\left( \mm{L}_0^{\otimes_{\rm p} x} \otimes_{\rm p} \mm{L}_1 \otimes_{\rm p}  \mm{L}_0^{\otimes_{\rm p} (n-1-x)}\right)= 
\sum_{x=0}^{n-1} \hat{\cal S}^x\left(\tr_{\rm a}(\mm{L}_0^{\otimes_{\rm p}(n-1)}\otimes_{\rm p}\mm{L}_1)\right)\\
& =& \sum_{x=0}^{n-1} \hat{\cal S}^x\left(\bra{0}\mm{L}_0^{\otimes_{\rm p}(n-1)}\otimes_{\rm p}\mm{L}_1\ket{0}\right) +
\sum_{k=1}^{m-1}\sum_{x=0}^{n-1} \hat{\cal S}^x\left(\bra{k}\mm{L}_0^{\otimes_{\rm p}(n-1)}\otimes_{\rm p}\mm{L}_1\ket{k}\right). \quad
\end{eqnarray*}
Using expressions (\ref{qr2}) and (\ref{p}), the first and the second term clearly correspond to the respective terms on the RHS of expression \eref{lem}. Note the cancellation
of the on-site magnetization terms in the final expression for the first term of \eref{lem}.\qed
\end{pf}
See Fig.~\ref{Figure} and the corresponding caption for an intuitive picture.

\begin{deff}
The periodic-shift invariant sequence of operator sums $Y_n$ written in the form (\ref{lem}) is {\em quasilocal} if there 
exist positive constants $\gamma,\gamma',\xi > 0$, 
\begin{equation}
\| q_r \|_{\rm HS} \le \gamma e^{-\xi r}, 
\quad 
{\rm and} 
\quad 
\| p_n \|_{\rm HS} \le \gamma' e^{-\xi n}.
\label{ql}
\end{equation} 
\end{deff}
Again, quasilocality of periodic operator sums implies {\em pseudolocality} in the sense of Definition 2, i.e., $\exists K>0$, such that $\| Y_n\|^2_{\rm HS} \le K n$.

We shall proceed to show in the following section that both operator sequences $\{q_r\}$ and $\{p_n\}$ are exponentially decreasing in Hilbert-Schmidt norm,
i.e., that mPNTO $Y_n(\varphi)$ is quasilocal, for an appropriate domain of $\varphi$.

\section{Proof of quasilocality}

\label{proof}

Now we are in position to state and prove the main result of the paper:

\newdefinition{thm}{Theorem}
\begin{thm}
For a dense set of easy-plane anisotropies $\eta = \pi l/m$, for coprime $l,m\in\ZZ^+$, $m \ne 0$, $l \le m$, translationally invariant operator sequences $Z_n(\varphi)$, for open boundaries as defined in (\ref{Zdef}), and $Y_n(\varphi)$, for periodic boundary conditions as defined in (\ref{Ydef}), are {\em quasilocal}, holomorphic operator-valued functions on the corresponding open vertical strips ${\cal D}_{m} = \{ \varphi; |\Re \varphi - \frac{\pi}{2}| < \frac{\pi}{2m} \}$.
\end{thm}
\begin{pf}
The key tool of our constructive proof will be a $(m-1)\times (m-1)$ {\em transfer matrix} defined on a reduced auxiliary space ${\cal H}'_{\rm a}={\rm lsp}\{\ket{k};k=1,\ldots,m-1\}$:
\begin{eqnarray}
 \mm{T}(\varphi,\varphi') \!\!&=&\!\!\!\sum_{k=1}^{m-1} (c^2_k +\cot\varphi\cot\varphi' s^2_k)\ket{k}\!\bra{k} + \sum_{k=1}^{m-2}\frac{|s_{k}s_{k+1}|}{2\sin\varphi\sin\varphi'}\left(\ket{k}\!\bra{k\!+\!1} + \ket{k\!+\!1}\!\bra{k}\right), \nonumber\\
&&{\rm where}\quad c_k := \cos(\pi l k/m), \quad s_k := \sin(\pi l k/m),  \label{TT}
\end{eqnarray}
\end{pf}
by which one facilitates computation of Hilbert-Schmidt products of local densities
\begin{equation}
\kappa_r(\varphi,\varphi'):=\frac{1}{2^{r}}\tr\left( q^T_r(\varphi) q_r(\varphi')\right) = \frac{1}{4} \bra{1} \mm{T}(\varphi,\varphi')^{r-2}\ket{1}, \quad {\rm r \ge 2}.
\label{rr}
\end{equation}
In order to demonstrate Eq.~(\ref{TT}) let us first list explicitly the Lax components (the transition operators of Fig.~\ref{Figure})
\begin{eqnarray}
\mm{L}_0^0 &=& \sum_{k=0}^{m-1} c_k \ket{k}\bra{k},\;\;\,\qquad\qquad\quad \mm{L}^0_1 = \eta \sum_{k=1}^{m-1} s_k \ket{k}\bra{k}, \nonumber\\
\mm{L}_0^\z &=& -\cot\varphi \sum_{k=1}^{m-1} s_k \ket{k}\bra{k},\;\;\,\quad\quad \mm{L}_1^\z = \eta\cot\varphi \sum_{k=0}^{m-1} c_k \ket{k}\bra{k},\nonumber\\
\mm{L}_0^+ &=& -\csc\varphi \sum_{k=1}^{m-2} s_k \ket{k+1}\bra{k},\quad \mm{L}_1^+ = 2\eta\csc\varphi\sum_{k=0}^{m-2} c_k \ket{k+1}\bra{k},\nonumber\\
\mm{L}_0^- &=& \csc\varphi \sum_{k=0}^{m-2} s_{k+1} \ket{k}\bra{k+1},\quad \mm{L}_1^- = 0.\label{L01}
\end{eqnarray}
Then we apply the representation (\ref{qr2}) to LHS of (\ref{rr}), together with $\bra{0}\mm{L}^-_0 = s_1\csc\varphi \bra{1}$, and
$\mm{L}^+_1\ket{0} = 2\eta\csc\varphi \ket{1}$, and write the remaining multiple sum over $\alpha_2,\ldots,\alpha_{r-1}$ in the Hilbert-Schmidt product as a power $r-2$ of a matrix over 
${\cal H}_{\rm a}\otimes {\cal H}_{\rm a}$, namely $2^{-r}\tr\left( q^T_r(\varphi') q_r(\varphi)\right) = \frac{1}{4}\bra{1}\otimes\bra{1}\TT^{r-2}\ket{1}\otimes\ket{1}$ with 
$\TT = \frac{1}{2}\sum_{\alpha\in{\cal J}} \mm{L}_0^\alpha(\varphi) \otimes \mm{L}_0^\alpha(\varphi') (\tr (\sigma^\alpha)^T\!\sigma^\alpha)$. Since $\TT$ preserves the subspace of `diagonal' vectors 
${\cal H}_\dd={\rm lsp}\{ \ket{k}\otimes\ket{k}; k=1,\ldots,m-1\}$, $\TT{\cal H}_\dd \subseteq {\cal H}_\dd$, we identify ${\cal H}_\dd$ with ${\cal H}'_{\rm a}$. More precisely, identification of basis states $\ket{k}\otimes\ket{k} \leftrightarrow |s_k| \ket{k}$, 
$ \bra{k}\otimes\bra{k} \leftrightarrow |s_k|^{-1} \bra{k}$ makes
 $\TT$ reading exactly as\footnote{The rescaling of basis, preserving bra-ket orthonormality, is needed to make $\mm{T}(\varphi,\varphi')$ (conveniently) symmetric.} expression (\ref{TT}).
We can use the same transfer matrix to write the Hilbert-Schmidt product of the remainders (\ref{p}):
\begin{equation}
\frac{1}{2^n}\tr\left( p^T_n(\varphi) p_n(\varphi')\right) = \tr\left\{ \mm{T}(\varphi,\varphi')^{n-1} \mm{V}(\varphi,\varphi')\right \},
\label{pp}
\end{equation}
where the vertex matrix $\mm{V}$ is obtained, similarly as before, by projection onto ${\cal H}'_{\rm a}$ of the following transfer matrix
$\VV = \frac{1}{2}\sum_{\alpha\in{\cal J}} \mm{L}_1^\alpha(\varphi) \otimes \mm{L}_1^\alpha(\varphi') (\tr (\sigma^\alpha)^T\!\sigma^\alpha)$,
\begin{equation}
\mm{V}(\varphi,\varphi') = \sum_{k=1}^{m-1}\eta^2(s^2_k +\cot\varphi\cot\varphi' c^2_k)\ket{k}\!\bra{k} + \sum_{k=1}^{m-2}\frac{2\eta^2 c_k^2 |s_{k+1}|}{|s_k|\sin\varphi\sin\varphi'} \ket{k\!+\!1}\bra{k}.
\label{V}
\end{equation}
Note that, with definition (\ref{pp}) and explicit representation of transition operators (\ref{L01}), one immediately sees that the state $\ket{0}$ is {\em never visited}, justifying projection ${\cal H}_{\rm a}\to{\cal H}'_{\rm a}$ in representation (\ref{pp}).
Then we proceed in the following steps:
\smallskip

\noindent
(i) The operators $q_r(\varphi), p_n(\varphi)$ are all holomorphic matrix-valued functions of $\varphi$
\begin{equation}
(q_r(\varphi))^\dagger = q^T_r(\bar{\varphi}),\qquad
(p_n(\varphi))^\dagger = p^T_n(\bar{\varphi}),
\end{equation}
which is following from definitions (\ref{qr2},\ref{p}) with explicit $\varphi-$dependences given in (\ref{L01}), and consequently,
\begin{eqnarray}
\| q_r(\varphi)\|^2_{\rm HS} &=& \frac{1}{4}\bra{1} \mm{T}(\bar{\varphi},{\varphi})^{r-2}\ket{1}, \quad {\rm r \ge 2}, \nonumber \\
\| p_n(\varphi)\|^2_{\rm HS} &=& \tr\left\{\mm{T}(\bar{\varphi},{\varphi})^{n-1} \mm{V}(\bar{\varphi},{\varphi})\right\}. \label{pqHS}
\end{eqnarray}
(ii) Next we will show that if $\varphi\in{\cal D}_m$ then the matrix $\mm{T}\equiv\mm{T}(\bar{\varphi},{\varphi})$ is strictly {\em contracting}, i.e., its eigenvalues $\tau_j(\varphi)$, if properly ordered, satisfy
$1 > |\tau_1| \ge |\tau_2| \ge \ldots \ge |\tau_{m-1}|$. Let us write ${\rm Re\,}\varphi=\frac{\pi}{2}+u$. Defining a positive diagonal matrix 
\begin{equation}
\mm{D}=\sum_{k=1}^{m-1} |s_k| \ket{k}\bra{k},
\end{equation}
and a tridiagonal Toeplitz matrix 
\begin{equation}
\mm{A}=\cos(2u)\one-\mm{E},\quad{\rm where}\quad \mm{E}=\frac{1}{2}\sum_{k=1}^{m-2}\left(\ket{k}\bra{k+1}+\ket{k+1}\bra{k}\right),
\end{equation} we have
\begin{equation}
\one - \mm{T} = |\sin\varphi|^{-2} \mm{D}\mm{A}\mm{D}.
\label{decomp}
\end{equation} 
All matrix elements of $\mm{T}$ are real and non-negative so the leading eigenvalue should be positive $\tau_1>0$, and $\mm{T}$ is contracting if $\one - \mm{T} > 0$. 
This is equivalent to condition $\mm{A} > 0$, or equivalently, $\mm{E} < \cos(2u)\one$, which holds if $|u| < \frac{\pi}{2m}$, i.e., $\varphi \in {\cal D}_m $.
\small skip

\noindent
(iii) The matrix $\mm{T}$ is real and symmetric and can be diagonalized $\mm{T}=\mm{O}\,{\rm diag}\{ \tau_j\} \mm{O}^T$, which, when applied to (\ref{pqHS}), yields quasilocality \eref{ql}, with 
\begin{equation}
\xi(\varphi) = -\frac{1}{2}\log \tau_1(\varphi) > 0,
\end{equation} 
and prefactors $\gamma,\gamma' > 0$ which in general depend on $\varphi$ as well. \qed

From Hilbert-Schmidt orthogonality of Pauli matrices and definitions (\ref{qr},\ref{rr}) the following useful orthogonality identities follow, for $x,x'\in\ZZ_n$ and $2\le r,r'\le n$:
\begin{eqnarray}
\frac{1}{2^n}\tr\left\{ \hat{\cal S}^{x}(\one_{2^{n-r}}\otimes q^T_{r}(\varphi)) \hat{\cal S}^{x'}\!(\one_{2^{n-r'}}\otimes q_r(\varphi'))\right\}
&=& \delta_{r,r'}\delta_{x,x'}\kappa_r(\varphi,\varphi'), \label{qtq} \\
\frac{1}{2^n}\tr\left\{ \hat{\cal S}^{x}(\one_{2^{n-r}}\otimes q_{r}(\varphi)) \hat{\cal S}^{x'}\!(\one_{2^{n-r'}}\otimes q_r(\varphi'))\right\}
&=& 0. \label{qq}
\end{eqnarray}
These immediately imply pseudolocality of operators $Z_n(\varphi)$ (\ref{Zdef}), and $Y_n(\varphi)$ (\ref{Ydef})
where Eqs. (\ref{extensivity},\ref{CS},\ref{pqHS}) are used to manipulate and finally estimate the effect of the remainder $p_n(\varphi)$:
\begin{eqnarray}
\| Z_n(\varphi)\|^2_{\rm HS} &=& n \sum_{r=2}^n \left(1-\frac{r-1}{n}\right)\|q_r \|^2_{\rm HS} \le n \gamma^2 \sum_{r=2}^n e^{-2\xi r} < n\frac{\gamma^2}{1-e^{-2\xi}},\\
\| Y_n(\varphi)\|^2_{\rm HS} &=& n \sum_{r=2}^n \|q_r\|^2_{\rm HS} + 2\Re \sum_{x=0}^{n-1}\sum_{r=2}^{n}\frac{1}{2^{n}}\tr\left\{p^\dagger_n \hat{\cal S}^x(\one_{2^{n-r}}\otimes q_r)\right\} + \| p_n \|^2_{\rm HS} \nonumber\\
&\le& n \gamma^2 \sum_{r=2}^n e^{-2\xi r} + 2n\gamma' \gamma e^{-\xi n}\sum_{r=2}^n e^{-\xi r} + {\gamma'}^2 e^{-2\xi n} \nonumber \\
& < & n \left(\frac{\gamma^2}{1-e^{-2\xi}} + \frac{2\gamma\gamma'}{1-e^{-2\xi}}e^{-\xi n}\right) + {\gamma'}^2 e^{-2\xi n}.
\label{end}
\end{eqnarray}
Clearly, the end expression (\ref{end}) can be estimated by $K n$ for a suitable $K>0$.

\section{Spin flip parity}

\label{spinflip}

The $XXZ$ model can be characterized in terms of a particularly important $\ZZ_2$ symmetry, namely the spin flip parity. 
We shall here focus only on periodic boundary conditions even though the same discussion applies to open boundaries as well. 
Defining the parity operator as
\begin{equation}
P = (\sigma^\x)^{\otimes n} = P^\dagger = P^{-1},
\end{equation}
one realizes that both, the hamiltonian $H_{\rm pbc}$ as well as the whole family of transfer operators 
in fundamental representation $V_n(\varphi,1/2)$ (as well as in any other finite dimensional irrep.) and consequently, the standard family of local conserved operators $Q^{(j)}_n,j=1\ldots,n-1$, commute with it
\begin{equation}
[H_{\rm pbc},P] = 0,\quad [Q^{(j)}_n,P] = 0,\quad [V_n(\varphi,s),P] = 0\quad{\rm for}\quad 2s\in\ZZ^+.
\end{equation}
The latter directly follows from spin flip symmetry (\ref{aparity}) in the auxiliary space for half-integer auxiliary spin. 
On the other hand, some important {\em nonequilibrium} physical observables, like the spin current operator
\begin{equation}
J_n = \ii \sum_{x=0}^{n-1} (\sigma^+_x \sigma^-_{x+1} - \sigma^-_x \sigma^+_{x+1}),
\label{sc}
\end{equation} 
or magnetization, anticommute
\begin{equation}
J_n P = - P J_n, \quad M^\z_n P = -P M^\z_n.
\end{equation}
As a consequence the expectation value any observable $A$ anticommuting with $P$, $A P = - P A$, in equilibrium state should vanish
since $\tr(e^{-\beta H_{\rm pbc}} A) = \tr(e^{-\beta H_{\rm pbc}}P^2 A) = -\tr(e^{-\beta H_{\rm pbc}} P A P) = -\tr(P e^{-\beta H_{\rm pbc}} P A) = -\tr(e^{-\beta H_{\rm pbc}} A)$.
We shall declare an operator $A$ for which $A P = P A$, or $A P = -P A$, to be of {\em even} ($\nu=1$), or {\em odd} ($\nu=-1$) parity, respectively. 
Clearly, the product of an operator of parity $\nu$ and an operator of parity $\nu'$ is an operator of parity $\nu\nu'$.
Therefore, negative parity observables are {\em invisible} for the entire standard machinery of (algebraic) Bethe ansatz \cite{korepin}.

Let us now show that the non-Hermitian transfer operators behave nontrivially under $P$. Straightforward inspection from the definitions reveals the following $PT-$like \cite{prl12} symmetry
\begin{equation}
P V_n(\varphi,s) P = V^T_n(\pi-\varphi,s),\qquad P Y_n(\varphi,s) P = Y^T_n(\pi-\varphi),
\end{equation}
and similarly with $W_n$ and $Z_n$ for open boundaries.
Note that the quasilocality domain ${\cal D}_m$ is symmetric under $\varphi \to \pi-\varphi$.
It is therefore useful to decompose the quasilocal conserved operators into even and odd components, $Y_n(\varphi) = Y^+_n(\varphi) + Y^-_n(\varphi)$,
\begin{equation}
Y_n^\pm(\varphi) := \frac{1}{2}(Y_n(\varphi) \pm P Y_n(\varphi) P) = \frac{1}{2}(Y_n(\varphi) \pm Y_n^T(\pi-\varphi)) 
\end{equation} 
satisfying $Y^\pm_n(\varphi) P = \pm P Y^\pm_n(\varphi)$.
$Y^-_n(\varphi)$ is thus expected to play particularly important role in nonequilibrium applications (see e.g. section \ref{spindrude}, or Ref.~\cite{mpp14}).

\section{Twisted boundary conditions}

\label{twisted}

Here we describe a simple modification of (quasilocal) non-Hermitian transfer operators which enables their exact commutation with the Hamiltonian $H_\phi$ (\ref{Hphi}) with twisted boundary condition. 
The key will the the following diagonal gauge matrix $\exp(\ii\phi\mm{S}^\z_s)$ which produces a fixed flux-phase upon commutation with spin raising/lowe\-ring operators in 
$m-$dimensional representation ${\cal V}_s$ (following from algebra (\ref{Uq}))
\begin{equation}
\exp(\ii\phi\mm{S}^\z_s)\,\mm{S}^\pm_s \exp(-\ii\phi\mm{S}^\z_s) = e^{\pm\ii \phi} \mm{S}^\pm_s.
\end{equation}
As a result, we have $U(1)$ symmetry of the Lax operator over ${\cal H}_{\rm a}\otimes{\cal H}_{\rm p}={\cal V}_s\otimes {\cal V}_{1/2}$
\begin{equation}
\exp(\ii\phi\mm{S}^\z_s)\,\mm{L}(\varphi,s) \exp(-\ii\phi\mm{S}^\z_s) =
\pmatrix{e^{-\ii\phi/2} & 0 \cr
0 & e^{\ii\phi/2}}\mm{L}(\varphi,s)
\pmatrix{e^{\ii\phi/2} & 0 \cr
0 & e^{-\ii\phi/2}}.
\label{U1}
\end{equation}
And as a further result of that, and of YBE over 
 ${\cal V}_{s}\otimes {\cal V}_{s'}\otimes {\cal V}_{1/2}$ and ${\cal V}_{1/2}\otimes {\cal V}_{1/2}\otimes {\cal V}_{s}$, one finds that the following {\em twisted non-Hermitian transfer operator} TNTO (see Ref.~\cite{bazhanov} for a related concept in the isotropic $XXX$ model) 
\begin{equation}
V_n(\varphi,s;\phi) = \tr_{\rm a}\left\{ \mm{L}(\varphi,s)^{\otimes_{\rm p} n}\exp(-\ii\phi\mm{S}^\z_s)\right\}.
\label{Vtdef}
\end{equation}
commutes with all the members of its family as well as with the Hamiltonian $H_\phi$
\begin{equation}
[V_n(\varphi,s;\phi),V_n(\varphi',s';\phi)] = 0,\quad [H_\phi,V_n(\varphi,s;\phi)] = 0, \quad \forall s,s',\varphi,\varphi'.
\end{equation}
Similarly as in purely periodic case we define the modified twisted non-Hermitian transfer operators (mTNTO)
\begin{eqnarray}
Y_n(\varphi;\phi) &=&  \frac{1}{2(\sin\varphi)^{n-2}\eta\sin\eta}(\partial_s + \ii\phi) V_n(\varphi,s;\phi)\vert_{s=0}-\frac{\cos\varphi\sin\varphi}{2\sin\eta}M^\z_n
\label{Ytdef} \\
&=&  \frac{\sin^2\varphi}{2\eta\sin\eta}\tr_{\rm a}\left\{\bra{0}_{\rm b}\tilde{\mm{L}}(\varphi)^{\otimes_{\rm p} n}\mm{G}_\phi \ket{1}_{\rm b}\right\} - \frac{\cos\varphi\sin\varphi}{2\sin\eta}M^\z_n,
\end{eqnarray}
where $\mm{G}_\phi := \exp(-\ii\phi\mm{S}^\z_0) = {\rm diag}(1,e^{\ii\phi},e^{2\ii\phi}\ldots e^{(m-1)\ii\phi})$, acting as a {\em scalar} in physical space ${\cal H}_{\rm p}$ as well as on derivative anzilla
${\cal H}_{\rm b}$. The second term on the RHS of (\ref{Ytdef}) is subtracted in order to conveniently compensate for the operator which is obtained when the $s-$derivative hits the gauge matrix
$\exp(-\ii\phi\mm{S}^\z_s)$ noting that $\partial_s\mm{S}^\z_s|_{s=0}=\one$, while the last term is still there to compensate for the trivial component in the direction of total magnetization. As all the three terms are mutually commuting, we have
again
\begin{equation}
[Y_n(\varphi;\phi),Y_n(\varphi';\phi)]=0,\quad \forall\varphi,\varphi'.
\end{equation}
Using canonical transformation (\ref{canonical},\ref{Hphi2}) one can write $Y'_n(\varphi;\phi) = C_\phi Y_n(\varphi;\phi) C^\dagger_\phi$ and use $U(1)$ symmetry (\ref{U1}) to distribute
the gauging phase homogeneously
\begin{equation}
Y'_n(\varphi;\phi) =  \frac{\sin^2\varphi}{2\eta\sin\eta} \tr_{\rm a}\left\{\bra{0}_{\rm b}\left(\tilde{\mm{L}}(\varphi)\mm{G}_{\phi/n}\right)^{\otimes_{\rm p} n}\ket{1}_{\rm b}\right\}-
\frac{\cos\varphi\sin\varphi}{2\sin\eta}M^\z_n,
\end{equation}
so the resulting mTNTO becomes periodic-shift invariant
\begin{equation}
\hat{\cal S} Y'_n(\varphi;\phi) = Y'_n(\varphi;\phi).
\end{equation}
This means that $Y'_n$ can again be written as a periodic-shift invariant sum of local operators (\ref{lem}) according to lemma 1 with $\mm{L}^\alpha_{0,1}$ replaced by
$\mm{L}^\alpha_{0,1} \mm{G}_{\phi/n}$ in the expressions of local densities (\ref{qr2}), and remainders (\ref{p}), denoting them as $q_n(\varphi;\phi)$, and $p_n(\varphi;\phi)$,
respectively. As $\bra{0}\mm{G}_{\phi/n}=\bra{0}$, $\mm{G}_{\phi/n}\ket{0}=\ket{0}$ all the boundary transition conditions (\ref{bc}), crucial for establishing locality of separate terms,
remain intact.

Furthermore, also the quasilocality theorem 1 goes through without change in the presence of the flux $\phi$. In fact, since 
\begin{equation}
(q_r(\varphi,\phi))^\dagger=q_r^T(\bar{\varphi},-\phi),\quad
(p_n(\varphi,\phi))^\dagger=p_n^T(\bar{\varphi},-\phi), \quad \phi\in\RaR,
\end{equation}
one finds that Hilbert-Schmidt products (at fixed $\phi$) do not depend on $\phi$, as they can be facilitated with exactly the same transfer matrix (\ref{TT}) as a consequence of invariance of
diagonal space ${\cal H}_\dd$ where $\mm{G}_{-\phi}\otimes\mm{G}_\phi$ acts trivially:
\begin{eqnarray}
&&\frac{1}{2^{r}}\tr\left( q^T_r(\varphi;-\phi) q_r(\varphi';\phi)\right) = \kappa_r(\varphi,\varphi'), \nonumber \\
&&\| q_r(\varphi;\phi)\|_{\rm HS} = \| q_r(\varphi)\|_{\rm HS},\quad \| p_n(\varphi;\phi)\|_{\rm HS} = \| p_n(\varphi)\|_{\rm HS}.
\end{eqnarray}
As a further consequence, extensive quasilocal operator norms can only differ by exponentially small amount, since the mixed terms
$2^{-n}\tr\left\{p^T_n(\varphi;-\phi)\hat{\cal S}^x (\one_{2^{n-r}}\otimes q^T_r(\varphi';\phi))\right\}$ {\em will} in general depend on $\phi$,
\begin{equation}
\| Y_n(\varphi)\|_{\rm HS} - \| Y_n(\varphi;\phi) \|_{\rm HS} = {\cal O}(n e^{-\xi(\varphi) n}).
\end{equation}

\section{Applications: Drude weight bounds and time-averaged operators}

\label{appl}

\subsection{Inner products of quasilocal conservation laws}

Let us define an inner product which turns ${\rm End}({\cal H}^{\otimes n}_{\rm p})$ into a Hilbert space, namely $(A,B):=2^{-n}\tr A^\dagger B$. Then, by means of the results of section \ref{proof}, one can straightforwardly write the following, complete families of inner products
\begin{eqnarray}
\left(Z_n(\bar{\varphi}),Z_n(\varphi')\right)&=&\sum_{r=2}^n (n-r+1)\kappa_r(\varphi,\varphi')\nonumber \\
&=& n\sum_{r=2}^\infty \kappa_r(\varphi,\varphi') - \sum_{r=2}^\infty (r-1) \kappa_r(\varphi,\varphi') + {\cal O}(n e^{-\xi n}) \nonumber \\
&=& n K(\varphi,\varphi') + {\cal O}(1), \label{innerZ} \\
\left(Y_n(\bar{\varphi}),Y_n(\varphi')\right) &=& n \sum_{r=2}^\infty \kappa_r(\varphi,\varphi') + {\cal O}(n e^{-\xi n}) \nonumber \\
&=& n K(\varphi,\varphi') + {\cal O}(n e^{-\xi n}), \label{innerY}
\end{eqnarray}
while the inner products with the transposed quasi-local operators either vanish or are exponentially small [see Eqs.~(\ref{qtq},\ref{qq})]
\begin{equation}
\left(Z^T_n(\bar{\varphi}),Z_n(\varphi')\right) = 0,\quad \left(Y^T_n(\bar{\varphi}),Y_n(\varphi')\right) = {\cal O}(n e^{-\xi n}),
\label{innerYt}
\end{equation}
where $\xi = \min\{\xi(\varphi),\xi(\varphi')\} > 0$, for $\varphi,\varphi'\in {\cal D}_m$.
We note that inner products for open and periodic (or equivalently, twisted, see section \ref{twisted}) boundary cases have the same volume coefficient in the thermodynamic limit
\begin{equation}
K(\varphi,\varphi') = \sum_{r=2}^\infty \kappa_r(\varphi,\varphi') = \frac{1}{4}\bra{1}(\one-\mm{T}(\varphi,\varphi'))^{-1}\ket{1}, 
\label{gm}
\end{equation}
whereas we have a relative $\propto 1/n$ versus a much smaller $\propto e^{-\xi n}$ finite size correction in the respective cases.
To see that the geometric series (\ref{gm}), as well as the ${\cal O}(1)$ correction term in (\ref{innerZ}), converge $\forall\varphi,\varphi'\in{\cal D}_m$ one may simply use Cauchy-Schwartz inequality (\ref{CS}) to estimate each summand
\begin{equation}
|\kappa_r(\varphi,\varphi')| \le \|q_r(\varphi)\|_{\rm HS}\,\|q_r(\varphi')\|_{\rm HS} < \gamma(\varphi)\gamma(\varphi')e^{-(\xi(\varphi)+\xi(\varphi'))r}.
\end{equation}
In order to evaluate LHS of (\ref{gm}) we introduce $\ket{\psi}\in {\cal H}'_{\rm a}$ as a solution of a linear equation 
\begin{equation}
(\one - \mm{T}(\varphi,\varphi'))\ket{\psi} = \ket{1}.
\label{eqpsi}
\end{equation}
Furthermore, we generalize (\ref{decomp}) and rewrite the transfer matrix (\ref{TT}) for any pair of spectral variables in terms of
a convenient decomposition
\begin{equation}
\one - \mm{T}(\varphi,\varphi') = -(\csc\varphi\csc\varphi') \mm{D}\left\{\cos(\varphi+\varphi')\one + \mm{E}\right\} \mm{D}.
\end{equation}
Writing the components as 
\begin{equation}
\ket{\psi} = \sum_{j=1}^{m-1} \frac{|s_1|}{|s_j|} \psi_j \ket{j},
\end{equation} 
the equation (\ref{eqpsi}) then results in a second order difference equation
\begin{equation}
\psi_{j+1} + 2\cos(\varphi+\varphi')\psi_j + \psi_{j-1} = -\frac{2\sin\varphi \sin\varphi'}{s^2_1} \delta_{j,1}
\end{equation}
with boundary conditions $\psi_0 = \psi_m = 0$, having an explicit solution, for $j\ge 1$:
\begin{equation}
\psi_j = 2(-1)^j \frac{\sin\varphi\sin\varphi'}{s^2_1} \frac{\sin((m-j)(\varphi+\varphi'))}{\sin(m(\varphi+\varphi'))}.
\end{equation}
Noting that $\bra{1}(\one-\mm{T}(\varphi,\varphi'))^{-1}\ket{1}=\psi_1$ we finally obtain a compact expression
\begin{equation}
K(\varphi,\varphi') = -\frac{\sin\varphi\sin\varphi'}{2 s^2_1} \frac{\sin((m-1)(\varphi+\varphi'))}{\sin(m(\varphi+\varphi'))}.
\label{kernel}
\end{equation}

\subsection{Mazur-Suzuki bounds for a continuous family of conserved operators}

\label{mazursuzuki}

In preceding short papers \cite{prl11a,prl13} it has been shown how almost conserved quasi-local operators generate nontrivial lower bounds on the high temperature spin Drude weight.
Due to residual boundary terms the thermodynamic limit in such a case has to be carefully discussed, in particular it has to be taken prior to a long time limit. Due to non-quasilocality w.r.t. $C^*$ operator norm of the operators\footnote{This has been noted after the publication of Ref.~\cite{prl13}.} $Z_n(\varphi)$, the application of Lieb-Robinson bounds \cite{ip13} seems problematic for finite (non-infinite) temperatures.

However, one can avoid any sort of problems of this type (on the rigorous level) by considering the $XXZ$ chain with periodic (or twisted) boundary conditions with exactly conserved quasilocal operators $Y_n(\varphi)$. 
Let us consider the {\em dynamical susceptibility} for an arbitrary observable\footnote{In fact, for our analysis the operator $A$ does not have to be Hermitian.} $A\in {\rm End}({\cal H}^{\rm \otimes n}_{\rm p})$, defined in terms of a time-average as
\begin{equation}
D_n (A) := \frac{1}{2n} \omega_\beta(\bar{A}^2),\quad \bar{A} := \lim_{T\to\infty}\frac{1}{T}\int_0^T\dd t e^{\ii H_{\rm pbc} t} A e^{-\ii H_{\rm pbc} t},
\label{DA}
\end{equation}
where $\omega_\beta(\bullet) = \tr\left\{\bullet e^{-\beta H_{\rm pbc}}\right\}/\tr e^{-\beta H_{\rm pbc}}$. Suzuki's version \cite{suzuki} of the lower bound can be written rigorously for any fixed $n$, and thermodynamic limit $n\to\infty$ (if it exists) can be taken optionally at the end.
Existence of the limit of time integrals (\ref{DA}) in the definition of time-averaged observable $\bar{A}$ is not in question for any finite $n$, as it can be evaluated explicitly in the eigenbasis of $H_{\rm pbc}$. 

Let us discuss here how to facilitate a continuous holomorphic family of exactly conserved quasilocal observables $\{Y_n(\varphi);\varphi\in{\cal D}_m\subset\CC\}$ for explicit computation of a lower bound of $D(A)=\lim_{n\to\infty}D_n(A)$ in the high temperature regime $\beta \to 0$. 
Without loss of generality we may choose $A$ to have a fixed parity $\nu$, which means we need to consider only the corresponding family of conserved operators 
$Y^\nu_n(\varphi)$ while the others are all orthogonal $(A,Y^{-\nu}_n(\varphi)) = 0$.

We start by considering an arbitrary integrable but not necessarily a holomorphic function $f: {\cal D}_m\to\CC$ which defines an operator
\begin{equation}
B = \bar{A} - \int_{{\cal D}_m}\!\dd^2\varphi\,f(\varphi) Y^\nu_n(\varphi)
\label{B}
\end{equation} 
and write a trivial inequality\footnote{The reader should not confuse the operator-time-averaging notation with complex conjugation for non-operator-valued quantities.} 
\begin{eqnarray}
0 \le \frac{1}{2n}(B,B) &=& D_n(A) - \frac{1}{2n}\int_{{\cal D}_m}\!\!\!\dd^2\varphi f(\varphi)(A,Y^\nu_n(\varphi)) - \frac{1}{2n}\int_{{\cal D}_m}\!\!\!\dd^2\varphi \overline{f(\varphi)}(Y^\nu_n(\varphi),A)\nonumber \\
&+& \frac{1}{2n}\int_{{\cal D}_m}\!\!\!\dd^2\varphi\int_{{\cal D}_m}\!\!\!\dd^2\varphi' 
\overline{f(\varphi)}f(\varphi')\left(Y^\nu_n(\varphi),Y^\nu_n(\varphi')\right).
\end{eqnarray}
We used the conservation property (\ref{cons}), yielding $(e^{\ii H_{\rm pbc}t} A e^{-\ii H_{\rm pbc} t},Y^\nu_n(\varphi))=(A,Y^\nu_n(\varphi))$, implying 
$(\bar{A},Y^\nu_n(\varphi)) = (A,Y^\nu(\varphi))$. Let us define the components of $A$ along the conserved operators in terms of a {\em holomorphic} function
\begin{equation}
a(\varphi) := \lim_{n\to\infty}\frac{1}{n}(A,Y^\nu_n(\varphi)),
\end{equation}
assuming the limit $n\to\infty$ exists (this question being trivial if $A$ is a translationally invariant sum of local operators). The limit in the last term exists as well, due to asymptotics 
(\ref{innerY},\ref{innerYt}),
yielding
\begin{equation}
\lim_{n\to\infty} \frac{1}{2n}(Y^\nu_n(\varphi),Y^\nu_n(\varphi')) = \frac{1}{4} K(\bar{\varphi},\varphi'),
\end{equation}
accounting for the $\varphi \to \pi - \varphi$ symmetry of the kernel (\ref{kernel}).
Therefore the limit $D(A)=\lim_{n\to\infty} D_n(A)$, if it exists, should satisfy the inequality
\begin{equation}
D(A) \ge F[f]:= \int_{{\cal D}_m}\!\!\!\dd^2\varphi\, {\rm Re}(a(\varphi)f(\varphi)) - \frac{1}{4} \int_{{\cal D}_m}\!\!\!\dd^2\varphi\int_{{\cal D}_m}\!\!\!\dd^2\varphi'\,K(\bar{\varphi},\varphi')\overline{f(\varphi)}f(\varphi')
\label{estimate}
\end{equation}
for any $f$.  Optimizing RHS by asking the linear variation of the functional to vanish for any small complex variation $\delta\!f$ of the function, 
\begin{equation}
\delta F[f] = {\rm Re}\int\!\dd^2\varphi\,\overline{\delta\!f(\varphi)}\left\{ \overline{a(\varphi)} - \frac{1}{2}\int\!\dd^2\varphi' K(\bar{\varphi},\varphi')f(\varphi')\right\} = 0,
\end{equation} 
where the symmetry of the kernel $K(\varphi,\varphi')=K(\varphi',\varphi)$ and the fact that it is holomorphic in both variables has been used, results in the complex Fredholm equation of the first kind for the unknown function $f$ (noting that $\overline{{\cal D}_m} = {\cal D}_m$):
\begin{equation}
\frac{1}{2}\int_{{\cal D}_m}\!\dd^2\varphi' K(\varphi,\varphi')f(\varphi') = \overline{a(\bar{\varphi})}.
 \label{Fredholm}
\end{equation}
The solution of the above equation can be plugged back to the estimate (\ref{estimate}) to yield the final Mazur-Suzuki lower bound
\begin{equation}
D(A) \ge \frac{1}{2}{\rm Re}\!\int_{{\cal D}_m}\!\!\!\dd^2\varphi\, a(\varphi)f(\varphi). 
\label{DAb}
\end{equation}

\subsection{Spin Drude weight}

\label{spindrude}

The recipe can be immediately demonstrated on the important example of the high temperature spin Drude weight $D_{\rm spin} = \beta D_J$, taking a spin current $A=J_n$ (\ref{sc}) and the odd parity set $\{Y^-_n(\varphi)\}$, yielding a constant coefficient $a(\varphi) \equiv \ii/4$. One finds, quite remarkably, that the integral equation (\ref{Fredholm}) is in this case solved by a simple function
\begin{equation}
f(\varphi) = -\ii \frac{m s_1^2}{\pi} \frac{1}{|\sin\varphi|^4} .
\label{f}
\end{equation}
Another elementary integral then yields the lower bound \cite{prl13} $D_J \ge D_K/4$,
\begin{equation}
D_K = \frac{\sin^2(\pi l/m)}{\sin^2(\pi/m)}\left(1 - \frac{m}{2\pi}\sin\left(\frac{2\pi}{m}\right)\right).
\label{DK}
\end{equation}
It is remarkable that the lower bound (\ref{DK}) agrees exactly with the thermodynamic Bethe ansatz calculation \cite{zotos2} at the special -- isolated -- points of anisotropy $\eta = \pi/m$ corresponding to $q-$defor\-mation
at {\em primitive} roots of unity ($l=1$). Since Bethe ansatz calculation for other values of $l$ seems to be highly nontrivial and has not yet been performed, we can only conjecture that the bound (\ref{DK}) is in fact saturating the exact value of thermodynamic high temperature spin Drude weight.

\subsection{Operator time averaging}

It is clear that the susceptibility bound derived in subsection \ref{mazursuzuki} is saturating if and only if $(B,B) = 0$, i.e., $B=0$, meaning that (see Eq.~(\ref{B})) in such a case we have an explicit expansion of a time-averaged operator in terms of the quasi-local conserved operators $Y^\nu_n(\varphi)$
and the solution $f(\varphi)$ of the Fredholm equation (\ref{Fredholm}) 
\begin{equation}
\bar{A} = \int_{{\cal D}_m} \dd^2\varphi f(\varphi) Y^\nu_n(\varphi).
\end{equation}
Since $f$ has been calculated in the thermodynamic limit while time-average is defined for a finite $n$, we expect to have corrections which are,
in Hilbert-Schmidt norm, exponentially small in $n$. Note that in case $\nu=-1$ one should subtract the trivial component in the direction of magnetization $M^\z$ (namely, take such $A$ that $(A,M^\z)=0$), since it has been subtracted from the quasilocal conserved operators as well.
Writing 
\begin{equation}
\bar{A}=\frac{1}{2}(\bar{A}'+\nu P \bar{A}' P)
\end{equation} 
with $\bar{A}':=\int_{{\cal D}_m} \dd^2\varphi f(\varphi) Y_n(\varphi)$ one can then write an explicit expression for time-averaged operator in terms of sums of local operators
\begin{equation}
\bar{A}' = \sum_{x=0}^{n-1}\sum_{r=2}^n \hat{\cal S}^x(\one_{2^{n-r}}\otimes a_r) + {\cal O}(e^{-c n}),\quad c > 0,
\end{equation}
where $a_r\in {\rm End}({\cal H}_{\rm p}^{\otimes r})$ are densities of time-averaged operator which read
\begin{equation}
a_r = \int_{{\cal D}_m} \dd^2\varphi f(\varphi) q_r(\varphi),
\end{equation}
and can be expressed in terms of Pauli operators using explicit MPO expression for the densities $q_r$ (\ref{qr2}).
Defining (spectral) parameter-independent Lax operator components restricted to subspace ${\cal H}'_{\rm a}$, $\mm{B}^\alpha \in {\rm End}({\cal H}'_{\rm a})$, via
\begin{equation}
\mm{L}^0_0(\varphi)|_{{\cal H}'_{\rm a}}=:\mm{B}^0,\quad \mm{L}^\z_0(\varphi)|_{{\cal H}'_{\rm a}}=:\mm{B}^\z \cot\varphi,\quad \mm{L}^\pm_0(\varphi)|_{{\cal H}'_{\rm a}}=:\mm{B}^\pm \csc\varphi, 
\end{equation}
where explicit (tridiagonal) matrix representation can be read directly from (\ref{L01}), and noting two other facts: (i) components $\alpha=+$ and $\alpha=-$ always come in pairs so the final amplitude in each term of $q_r(\varphi)$ is an even order monomial in $\csc\varphi$, and (ii) 
$\csc^2\varphi = 1 + \cot^2 \varphi$, we write
\begin{equation}
a_2 = a^{\{\}}_2 \sigma^- \otimes \sigma^+,\quad
a_r = \sum_{s_{2}\ldots \alpha_{r-1}\in {\cal J}} a^{\alpha_2\ldots \alpha_{r-1}}_r \sigma^- \otimes \sigma^{\alpha_2} \otimes \cdots \sigma^{\alpha_{r-1}}\otimes \sigma^{+},\quad r > 2,
\end{equation}
where $a^{\alpha_2\ldots\alpha_{r-1}}_r$ are coefficients given as
\begin{eqnarray}
a^{\{\}}_2 &=& \int_{{\cal D}_m}\dd^2\!\varphi\,f(\varphi), \\
a^{\alpha_2\ldots\alpha_{r-1}}_r &=& 
\bra{1}\mm{B}^{\alpha_2}\cdots\mm{B}^{\alpha_{r-1}}\ket{1}\int_{{\cal D}_m}\!\dd^2\varphi\, (1+ \cot^2\varphi)^{\#_+\{\alpha_i\}} (\cot\varphi)^{\#_\z\{\alpha_i\}}.
\label{BB}
\end{eqnarray}
Here $\#_\alpha \{\alpha_i\}$ denotes the number of occurrences of index $\alpha$ in the list $\{\alpha_i\}\equiv \alpha_2\ldots \alpha_{r-1}$.
With some combinatorics the latter integral can be expressed in terms of pure monomials
\begin{equation}
I_k = \int_{{\cal D}_m}\!\dd^2\varphi\,f(\varphi) (\cot\varphi)^{2k},\quad k\in \ZZ^+
\label{Ik}
\end{equation}
while noting that the corresponding integrals with {\em odd} monomials vanish due to reflection symmetry $\varphi \to \pi - \varphi$ of the domain ${\cal D}_m$, i.e.,
$I_{k+1/2}\equiv 0$,
\begin{equation}
\int_{{\cal D}_m}\!\dd^2\varphi\, (1+ \cot^2\varphi)^{\#_+\{\alpha_i\}} (\cot\varphi)^{\#_\z\{\alpha_i\}} = \sum_{j=0}^{\#_+\{\alpha_i\}}
{\#_+\{\alpha_i\}\choose j} I_{j+\frac{1}{2}\#_\z\{\alpha_i\}}.
\label{csum}
\end{equation}

\subsection{Time-averaged spin current}

A straightforward explicit calculation of the time-averaged spin-current (\ref{sc}) (or particle current in the related interacting spinless fermion model) $\bar{J}$ has recently been reported in \cite{mpp14}. In this case, the integrals (\ref{Ik}) can be explicitly calculated due to simplicity of the function $f$ $\label{f}$ and the fact that under 
{\em conformal} transformation $z=\cot\varphi$, the integrals (\ref{Ik}) map to simple algebraic monomials 
\begin{equation}
I_k=-\ii \frac{m s_1^2}{\pi} \int_{{{\cal D}'_m}}\!\dd^2 z\, z^{2k},
\label{Ikz}
\end{equation} 
whereas $1/|\sin \varphi |^4 = |\dd z/\dd\varphi|^2$ from $f(\varphi)$ is just the Jacobian of the conformal mapping\footnote{
It is perhaps worth remarking that $z=\cot\varphi$ could be used as a spectral variable all the way through our analysis, with the convenience that Lax operator components $\mm{L}^\alpha_{0,1}$ could be written as {\em linear} functions of $z$.} which maps the domain ${\cal D}_m \to{\cal D}'_m$ to an intersection of two disks of equal radii 
$\csc(\pi/m)$ and centers at $\pm \cot(\pi/m)$, intersecting under angle $\pi/m$ at the corners $\pm\ii$. An exercise in elementary analysis then yields simple expressions for the integrals (\ref{Ikz})
\begin{equation}
 I_k = \frac{\ii (\csc \frac{\pi}{m})^{2k}}{(2k+1)}\sum_{j=0}^{2k+1}(-1)^j {2k+1\choose j}\left(
 \mathrm{sinc}\left(\frac{\pi (j+1)}{m}\right)-\mathrm{sinc}\left(\frac{\pi(j-1)}{m}\right)\right)\left(\cos\frac{\pi}{m}\right)^{2k+1-j}.
 \label{Ikj}
\end{equation}
This concludes explicit representation of the time-averaged current $\bar{J}$ in terms of sums of local Pauli operators. 
Coefficient of each local term is efficiently computable in terms of a product of matrices (\ref{BB}) and simple combinatorial sums (\ref{csum},\ref{Ikj}),
whereas distinct nonvanishing terms can be completely enumerated by  means of the left tower of Fig.~\ref{Figure}.

\section{Conclusions}

In the present paper we have elaborated on a detailed derivation of quasi-local conservation laws for $XXZ$ spin-$1/2$ chain with periodic, or twisted boundary conditions.
Due to their intrinsically non-Hermitian character, these objects have access to the sector of observables with odd spin flip parity. Consequently, they have been shown to play an important role 
for understanding spin-transport features of the model. There are several interesting future challenges: (i) To extend Drude weight calculations/bounds to finite (non-infinite) temperatures, where analytical computation of Kubo-Mori inner product of quasi-local operators should be considerably more involved. (ii) Establish, on a rigorous level, if Mazur bound using our set of quasilocal operators is generally saturating or there could be still a gap, say 
for incommensurable anisotropies in the regime of easy-plane interactions. (iii) Develop analogous concepts (perhaps based on non-quasilocal higher spin $s-$derivatives of PNTOs around $s=0$) to systematically access {\em finite size corrections} to dynamical susceptibility bounds. (iv) To elaborate on such a construction in other integrable quantum models with the same trigonometric $R$-matrix,
like e.g. Sine-Gordon quantum field theory or its integrable discretizations.

\section*{Acknowledgements}
The work has been supported by grants P1-0044 and J1-5439 of Slovenian Research Agency.
The author thanks E. Ilievski, M. Mierzejewski and P. Prelov\v sek, for discussions and collaboration on related previous work, and acknowledges an inspiring communication with A. Kl\" umper.

\section*{Note added}
A closely related independent work \cite{pereira}, proposing essentially equivalent concepts, appeared on the public preprint repository just after the manuscript of the present work.

\end{document}